\documentclass[aps,pra,twocolumn,showpacs]{revtex4}
\usepackage{graphicx}\usepackage{amssymb}\usepackage[english]{babel}

\newcommand{\greatersim}{\mathrel{\mbox{\raisebox{0.45ex}{
\hspace{-0.25em}${>}$\hspace{-0.77em}\raisebox{-1.2ex}{${\sim}$}}}}}
\begin{document}
\title{Bose-Einstein condensation of 2D dipolar excitons: Quantum Monte Carlo
simulation}
\author{Yu.E.~Lozovik$^a$, I.L.~Kurbakov$^a$, G.E.~Astrakharchik$^b$, and
Magnus~Willander$^c$}
\affiliation{$^a$ Institute of Spectroscopy, Russian Academy of Sciences,
142190 Troitsk, Moscow region, Russia\\
$^b$ Departament de F\'isica i Enginyeria Nuclear, Campus Nord B4-B5,
Universitat Polit\`ecnica de Catalunya, E-08034 Barcelona, Spain\\
$^c$ Institute of Science and Technology (ITN), Link\"oping University,
SE-581 83, Link\"oping, T\"appan 6177, Sweden}
\date{\today}
\begin{abstract}
The Bose condensation of 2D dipolar excitons in quantum wells is numerically
studied by the diffusion Monte Carlo simulation method. The correlation,
microscopic, thermodynamic, and spectral characteristics are calculated. It is
shown that, in structures of coupled quantum wells, in which low-temperature
features of exciton luminescence have presently been observed, dipolar
excitons form a strongly correlated system. Their Bose condensation can
experimentally be achieved much easily than for ideal or weakly correlated
excitons.
\end{abstract}
\maketitle
\section{Introduction}\label{Intr} 
Two-dimensional (2D) dipolar excitons, in which electrons ($e$) and holes
($h$) are spatially separated, can be created in structures of coupled (CQWs)
quantum wells (QWs) \cite{LY} and single QW (SQW) in a strong external
transverse electric field \cite{js035197}. At low temperatures, $e$ and $h$ in
these systems reside in the lowest transverse (with respect to the QW plane)
quantized states. In this case, the dipole moments of excitons (in the ground
axially symmetric state) are all directed perpendicular to this plane. There
are two types of excitons in CQWs: direct excitons, whose electrons and holes
are in the same QW, and spatially indirect excitons, whose electrons
and holes are located in different QWs (see details in experimental studies
\cite{revM,revS,revT,revB,T,B,M,Snoke1,ss134037,rl970103,b7400409,transp}). In
the second case, exciton dipole moment is $d=eD$, where the quantity $D$ is
determined by the geometry and is equal to the characteristic interwell
distance (see Eq.~(\ref{D}) below). In single quantum wells, a dipolar exciton
can occur due to a strong transverse electric field, which induces a dipole
moment \cite{js035197,TSQW}. In both cases, unlike quasi-2D atomic gases
\cite{rl842551}, two 2D dipolar excitons cannot pass "over one another" in the
QW plane. Dipolar interaction of 2D excitons is theoretically studied in Ref.
\cite{2DDE}.

At low exciton densities, 2D dipolar excitons are far from each other, so
exchange effects between electrons (holes) of neighboring excitons are
suppressed by a dipole barrier, related to a exciton dipole-dipole repulsion, 
$$
P\sim\exp\left(-\int_{a_{ex}}^{a_{cl}}\sqrt{m(V(r)-2E_0/N)}dr\right).
$$
Here, $P$ is the probability for a 2D dipolar exciton to tunnel through the
dipole barrier; $m$ is the exciton mass; $E_0$ is the ground-state energy of
the excitons (their kinetic and dipolar-interaction energies); $N$ is the
exciton number; $V(r$) is the dipolar exciton potential; $a_{ex}$ is the
exciton diameter (which is of the order of the average distance $a_{eh}$
between $e$ and $h$ in the exciton); and $a_{cl}$ is the classical turning
point for the potential $V(r)$ ($V(a_{cl})=2E_0/N$). In this case, 2D dipolar
excitons can be considered as (composite) bosons. At high exciton densities,
at which the distance between neighboring excitons is smaller or of the order
of the exciton radius, the Fermi properties of the internal exciton structure
play a significant role. This leads to a transformation of a dipolar exciton
gas into an $e$-$h$ system \cite{LB0,prl100256402} (although, in the isotropic
system at $T=0$, the crossover from an exciton Bose condensate to a BCS state,
with the gap decreasing with an increase in the density, occurs \cite{LY}).

The possibility of the Bose condensation of 3D excitons was under
consideration long ago \cite{KK}. However, in a homogeneous infinite 2D
system, because of the ultraviolet divergence of the exciton momentum
distribution, the Bose condensation is possible only at $T=0$ \cite{r1580383}.
In the range of finite temperatures, only a quasi-long-range (power-law) order
\cite{PFCT} and a superfluid phase can occur, which are destroyed due to the
dissociation of pairs of oppositely directed vortices and formation of free
vortices at the Berezinskii-Kosterlitz-Thouless (BKT) transition point
\cite{Berezinskii,jc061181,jc071046,rl391201}. However, in {\it finite} 2D
systems (traps), the ultraviolet divergence of the momentum distribution is
cut off, so, the Bose condensate can exist \cite{rl842551,2DBEC}. This opens
opportunities for studying the Bose condensation of 2D dipolar excitons in
real experimental systems.

At present, the search for the Bose condensation of 2D dipolar excitons in
CQWs \cite{revM,revS,revT,revB} and in SQW has attracted much attention from
researchers. Interesting optical effects observed in studying the luminescence
of 2D dipolar excitons in CQWs at low temperatures point to significant
experimental progress in the investigation of the collective properties of 2D
dipolar excitons \cite{T,B,M,Snoke1,ss134037,rl970103,b7400409,transp}. The
observations of a narrow-beam luminescence from a 2D dipolar exciton Bose
condensate along a normal to the QW plane and of the off-diagonal long-range
order of Bose condensed 2D dipolar excitons in the coherence experiment have
recently been performed by Timofeev et al. in SQW in a transverse electric
field \cite{TSQW}.

Theoretically, for the collective state of 2D dipolar excitons in QWs,
superfluidity \cite{SF,SFB,LB}, Bose condensation \cite{LB0,LB,BEC} (in the
exciton regime), a BCS-like state \cite{LY,BCS} (in the $e$-$h$ system),
strong correlations \cite{js035197,MCLKAW}, crystallization \cite{js035197,%
cr,MCcr,rl980605,MCcrLKAW} and a mesoscopic supersolid phase \cite{SS},
kinetic and relaxation processes \cite{kinrel}, vortical \cite{V} and
quasi-Josephson \cite{QD} effects, optical properties \cite{opt} as well as
these in circular traps \cite{ring}, a superfluid transition (crossover) in
traps \cite{LKW}, as well as the effect of random fields \cite{SFB,rand}, were
studied.

The superfluidity of Bose condensed dipolar excitons was also studied in
two-layer $e$-$e$ systems in a strong magnetic field with the half-occupancy
of the Landau levels in each layer ($\nu=1/2+1/2$; see, e.g., the theory in
Ref.~\cite{2DEEt} and experiments in Ref.~\cite{2DEEe} and references
therein).

2D dipolar bosons were also studied in rarefied atomic gases \cite{dipat} and
polar molecules \cite{polmol}.

Among the above studies into the exciton Bose condensate, the model of an
ideal exciton gas and that of a weakly correlated exciton gas were
predominantly used as theoretical approaches. Unfortunately, these approaches
have a rather limited domain of applicability. Indeed, at low densities, the
repulsive dipole potential can be described by only one parameter, the length
$a_s$ of the isotropic $s$-scattering. Therefore, the properties of such a
potential should be universal, i.e., the same for all interaction potentials
having the same scattering length; in particular, the properties should be the
same as for the system of hard disks with the diameter $a_s$. However, the
latter system is known to be weakly correlated only in the ultrararefied case
\cite{HD}. Consequently, the model of weakly correlated excitons is valid only
for ultrararefied systems, whose critical temperature is extremely low. For
real experimentally observed densities, this model can give only a qualitative
description. Therefore, the quantitative theoretical investigation of the
2D QW dipolar exciton collective state requires a more accurate model and more
precise, desirably, ab initio, calculations. 

This work is devoted to the detailed microscopic investigation of 2D dipolar
excitons by means of the numerical simulation by the diffusion Monte Carlo
method. The main result of our simulation is the inference that, in the
experimentally observed low-temperature regime \cite{revT,revB,T,B,M}, 2D
dipolar excitons in CQWs form a strongly correlated system \cite{Butovring}.
We find the following evidences for this fact.

(i) The dimensionless adiabatic compressibility $\zeta=m^3/$
$\!\!(2\pi\hbar^2\chi)$ ($\chi$ is the corresponding dimensional quantity) and
the contribution $\zeta'=\mu_im/(2\pi\hbar^2n)$ to the chemical potential
$\mu_i$ related to the dipolar exciton-exciton interaction prove to be much
greater than unity, $\zeta,\zeta'\gg1$ (Here, $n$ is the total exciton density
in all the $g_{ex}$ spin degrees; $g_{ex}=4$ for GaAs). For weakly correlated
excitons, $\zeta,\zeta'\ll1$.

(ii) The density $n_0$ of the exciton Bose condensate (in all the $g_{ex}$
spin degrees) at $T=0$ proves to be two to four times smaller than the total
density $n$ (in the regime of weak correlations, $n_0\approx n$).

(iii) A clearly pronounced hump is observed both in the pair distribution
function and in the structure factor, which points to the presence of the
short-range order. In addition, at higher exciton densities, e.g., at
$n=5\cdot10^{10}$ cm${}^{-2}$, two humps are observed in the structure factor
and three in the pair distribution function.

(iv) The spectrum of excitons is very far from the Bogolyubov shape (although,
even at the boundary between the exciton and $e$-$h$ regimes, the roton
minimum is not yet reached under the experimental conditions of
Refs.~\cite{revT,revB,T,B,M}).

(v) The local superfluid exciton density at the superfluid transition
temperature, $n_l(T_c)$, is close to the total exciton density $n$ (whereas,
for weakly correlated systems, the quantity $n_l(T_c)$ is logarithmically
small compared to $n$ \cite{b3704936}) \cite{nl=n}.

(vi) The temperature of the superfluid transition in the corresponding
infinite system is only slightly less than the degeneracy temperature
$T_{deg}$, while the quasi-condensation temperature is 2-2.5 times higher than 
$T_{deg}$ (for the spin-depolarized exciton gas in all the $g_{ex}=4$ spin
degrees). In contrast, according to the theory of weakly correlated excitons,
these temperatures are logarithmically small compared to $T_{deg}$
\cite{b3704936}. 

(vii) The profile of the exciton Bose condensate in a large 2D harmonic trap
at $T=0$ appreciably differs from the Thomas-Fermi inverted parabola.

In some structures of CQWs \cite{Snoke1,ss134037} and SQW \cite{TSQW}, the
condensed gas of 2D dipolar excitons forms a system with intermediate
correlations ($\zeta,\zeta'\sim1$). In this case, the above evidences weaken
(evidence (iii) vanishes). However, in structures of CQWs and SQW with the
width of these wells being sufficiently large, at high densities of excitons,
their correlations prove to be so strong that a roton minimum appears (and
even the crystallization of excitons is possible; see, e.g.,
Refs.~\cite{rl980605,MCcrLKAW}). 

The paper is organized as follows. In Section \ref{measur}, the model is
discussed, the problem is formulated, and the relation with experiment is
considered. Also, the details of simulation are presented. Section
\ref{results} is devoted to a homogeneous exciton system. A detailed
analytical processing is constructed based on the ab initio numerical
simulation, with the microscopic parameters of the problem being calculated.
Using these results, in Section \ref{harm}, we
analytically investigate a large harmonic trap in the local density
approximation. The possibilities of experimental observation of strong
correlations in the system of 2D dipolar excitons in CQWs are discussed in
Section \ref{lum}. In section \ref{concl} we conclude the study.

\section{Model and calculation method}\label{measur}
Two 2D spatially separated excitons repel each other as dipoles,
$V(r)\propto1/r^3$, if the distance between them is much longer than the
effective spacing $D$ between $e$ and $h$ layers. In the exciton regime, the
distance between neighboring excitons in structures of Refs.~\cite{revT,revB,%
T,B,M,Snoke1,ss134037,rl970103,b7400409,transp,TSQW}, according to a rough
estimate (see details in \cite{b3206601}), exceeds $D$. Therefore, the model
of dipolar excitons is a good approximation for experimentally studied
systems. The suppression of the exchange interaction by the dipole barrier in
the exciton  regime makes it possible to consider 2D dipolar excitons as
structureless bosons \cite{LB0}.

As a result, at $r\gg D$, we arrive at the following Hamiltonian of the
homogeneous system of structureless 2D dipolar excitons:
\begin{equation}\label{H}
\hat H=\frac{\hbar^2}{mx_0^2}\left(\sum_i\frac{\bar\Delta_i}2
+\sum_{i<j}\frac1{\bar r_{ij}^3}\right).
\end{equation}
Here,
\begin{equation}\label{x0}
x_0=\frac{md^2}{4\pi\varepsilon\varepsilon_0\hbar^2}=
\frac{me^2D^2}{\varepsilon\hbar^2}>0
\end{equation}
is the parameter with the dimension of length for the exciton dipole-dipole
potential $V(r)$, 
$$
V(r)=\frac{\hbar^2}m\frac{x_0}{r^3},\;\;\;\;
\bar r_{ij}=\frac{|{\bf r}_i-{\bf r}_j|}{x_0},
$$
$$
\bar\Delta_i=x_0^2\Delta_i,\;\;\;\;
\Delta_i=\partial^2/\partial x_i^2+
\partial^2/\partial y_i^2+\partial^2/\partial z_i^2,
$$
$d=eD$ is the dipole moment of an exciton, $e>0$ is the hole charge,
$\varepsilon$ is the dielectric constant of the structure,
$\varepsilon_0=1/4\pi$ is the dielectric permittivity of vacuum, and 
\begin{equation}\label{D}
D=\left|\int_{-\infty}^{\infty}(|\psi_e(z)|^2-|\psi_h(z)|^2)zdz\right|
\end{equation}
is the effective distance between the $e$ and $h$ layers, where $\psi_e(z)$
and $\psi_h(z)$ are, respectively, the electron and the hole wave functions
along the $z$ axis.

We impose periodic boundary conditions. This, on the one hand, makes the
system finite and, on the other hand, imitates its homogeneity. In this case,
the momentum becomes discrete but remains good quantum number.

At the first stage, in calculations by the variational Monte Carlo method, we
specify the trial function of the ground state of excitons to be a
Bijl-Jastrow function 
\begin{equation}\label{psi}
\psi_T({\bf r}_1...{\bf r}_N)=\prod_{i<j} f(\bar r_{ij}),
\end{equation}
where, 
\begin{equation}\label{f}
f(\bar r)=\left\{\begin{array}{ll}
AK_0(2/\sqrt{\bar r}),\;&\bar r<\bar r_c,\\
B\exp(-C/\bar r-C/(\bar L-\bar r)),\;&\bar r_c\le\bar r<\bar L/2.
\end{array}\right.
\end{equation}
Here, $K_0(x)$ is the modified Bessel function of the second kind (the
Macdonald function) of the zeroth order, $\bar r=r/x_0$, and $\bar L=L/x_0$ is
the dimensionless size of the system $L$. The constants $A$, $B$, and $C$ are
chosen from the condition $f(\bar L/2)=1$ and continuity conditions of the
zeroth and first derivatives at $\bar r=\bar r_c$, and the joining point
$\bar r_c$ is used as the (only) variational parameter. We expect that the
trial wave function (\ref{psi}) is a good approximation for the wave function
of the ground state. When two excitons approach close each other, the
influence of remaining exciton pairs becomes small, and the function
$f(\bar r)$ corrresponds to the solution of the two-particle scattering
problem from the dipole-dipole potential. At short distances, function
(\ref{f}) is chosen such that it corresponds to the exact solution of the
scattering problem for zero energy. We also verified that the use of the
numerical solution to the scattering problem at finite energy as a
two-particle factor of $f(\bar r)$ in the Bijl-Jastrow function weakly varies
the variational energy of the system and almost does not affect the diffusion
value of the energy. The long-wavelength behavior of the function $f(\bar r)$
is determined by the collective properties and corresponds to the phonon
propagation in the 2D system: $f(\bar r)\propto e^{-const/\bar r}$
\cite{r1550088}. The second term is added to the exponent of the exponential
in Eq.~(\ref{f}) in order to satisfy the condition that the derivative at the
boundary of the box is zero: $f'(L/2)=0$. The function $f(\bar r)$ is shown in
Fig.~1. 

\begin{figure}[t]
\includegraphics[width=8.65cm,height=5.8cm]{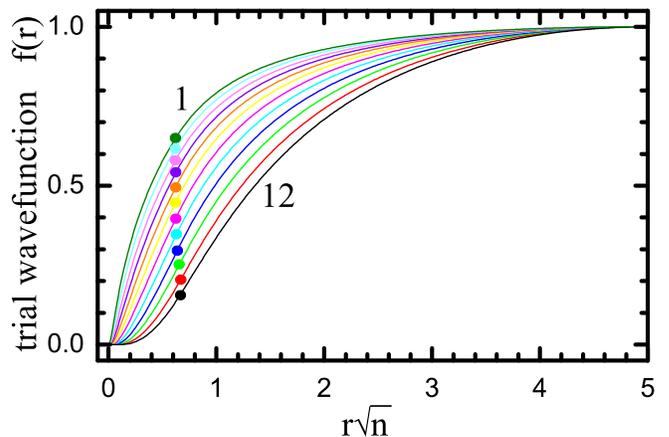}
\vskip -4mm
\caption{\small Trial wave function $f(r)$ at different exciton densities $n$.
Curve $i$ ($i=1,2...12$) corresponds to the density $nx_0^2=2^{i-9}$. Circles
show the point $r_c$ of joining of two asymptotics.}
\end{figure}

As is seen from Eqs.~(\ref{H}), (\ref{psi}), and (\ref{f}) the Hamiltonian and
the trial function are reduced to the dimensionless forms that are expressed,
respectively, in terms of the linear units $x_0$ and energy units
$\hbar^2/mx_0^2$. As a result, the dimensionless exciton density
$\bar n=nx_0^2$ is the only dimensionless parameter of the problem.

The quantity $1/x_0^2$ yields the 2D exciton density $n$, which corresponds to
the dimensionless density $\bar n=nx_0^2=1$ and which can be calculated by
Eq.~(\ref{x0}). Now, we will consider the range of variation of the
dimensionless density $\bar n$ in various real experimental systems.

(1) For the structure of GaAs/AlAs CQWs studied by Timofeev and colleagues
\cite{revT,T} ($D=13.6$ nm, $m=0.22m_0$ ($m_0=9.1083\cdot10^{-28}$ g is the
free electron mass), and $\varepsilon=12.5$), $1/x_0^2=2.64\cdot10^{10}$
cm${}^{-2}$.

(2) For the structure of GaAs/AlGaAs CQWs (Butov et al. \cite{revB,B},
$D=12.3$ nm, $m=0.22m_0$, and $\varepsilon=12.5$), $1/x_0^2=3.94\cdot10^{10}$
cm${}^{-2}$.

(3) For the structure of asymmetric GaAs/AlGaAs CQWs (Moskalenko et al.
\cite{M}, $D=14.1$ nm, $m=0.22m_0$, and $\varepsilon=12.5$),
$1/x_0^2=2.28\cdot10^{10}$ cm${}^{-2}$.

(4) For the structure of InGaAs/GaAs CQWs (Snoke et al. \cite{Snoke1,%
ss134037}, $D=10.5$ nm, $m=0.14m_0$, and $\varepsilon=12.5$),
$1/x_0^2=1.83\cdot10^{11}$ cm${}^{-2}$.

(5) For the structure of GaAs/AlGaAs CQWs (Snoke et al. \cite{rl970103,%
b7400409}, $D=15.5$ nm, $m=0.22m_0$, and $\varepsilon=12.5$),
$1/x_0^2=1.56\cdot10^{10}$ cm${}^{-2}$.

(6) For the structure of GaAs SQW (Timofeev et al. \cite{TSQW}, $D\approx6$
nm, $m=0.22m_0$, and $\varepsilon=12.5$), $n=1/x_0^2\sim7\cdot10^{11}$
cm${}^{-2}$, which corresponds to the dimensionless density $\bar n=1$.

Next, we have to know the range of variation of the 2D exciton density $n$ in
their experimentally achievable condensed state.

The upper limit of the density is the boundary between the strong and weak
coupling regimes (corresponding to the condensed exciton state and to the 
BCS-like $e$-$h$ state). For structure (1) \cite{revT,T} (with an average
electron-hole distance of $a_{eh}=17$ nm \cite{T}), the corresponding critical
density determined according to the rough estimate of Ref.~\cite{b3206601} is
\begin{equation}\label{nmax}
n_{max}=\frac{0.117g_{ex}}{\pi a_{eh}^2}\approx5.17\cdot10^{10}
\mbox{ cm}^{-2},\;\;\;\;\bar n_{max}\approx1.96
\end{equation}
(here, it is taken into account that, in GaAs, excitons are in the $g_{ex}=4$
spin degrees). 

Therefore, characteristic exciton densities that can be achieved in modern
experiments in QWs, at which low-temperature features in exciton luminescence 
are observed \cite{revT,revB,T,B,M}, lie in the range
$10^{10}\le n\le5\cdot10^{10}$ cm${}^{-2}$, i.e.,
$1/4\lesssim\bar n\lesssim4$.

However, in very wide (wider than 100-200 nm) SQWs based on GaAs, such high
dimensionless densities ($\bar n_{max}>290$ \cite{MCcrLKAW}) can be achieved
at which excitons crystallize \cite{rl980605}.

We performed simulation in the range of the dimensionless density
$1/256\le\bar n\le8$. For structure (5) \cite{rl970103,b7400409} (in which the
exciton correlations are stronger than in other structures), at the density
$\bar n=8$, the dimensional exciton density is high: $n=1.2\cdot10^{11}$
cm${}^{-2}$. In structure (6) \cite{TSQW} (where the exciton correlations are
weaker than in other structures), at the dimensionless density $\bar n=1/256$
(and at the corresponding dimensional density $n\approx2.7\cdot10^9$
cm${}^{-2}$), the superfluid crossover temperature of excitons in a trap of
about 100 $\mu$m in size is low: 0.12 K (see \cite{LKW} and Section
\ref{outnl}).

We should note that, in a sufficiently weak electric field, very small dipole
moments $eD$ and very low dimensionless densities $\bar n=nx_0^2\propto D^4$
(see Eq.~(\ref{x0})) can be realized in SQWs. However, we are not interested
in this case. Indeed, at $D\to0$, the contribution of the van der Waals
potential to the exciton scattering length $a_s$ ($a_s^{WDW}\sim a_{eh}$)
considerably exceeds the contribution from the dipole potential to this
parameter ($a_s^D=3.17222x_0\propto D^2\to0$). As a result, excitons
efficiently loose the property of dipolarity.

Nevertheless, in GaAs-based SQWs ($m=0.22m_0$, $\varepsilon=12.5$,
$a_{eh}\approx12$ nm), in a sufficiently weak electric field, the
dimensionless density of a rarefied 2D dipolar exciton gas with a superfluid
crossover temperature of about 0.1 K and with a dipole scattering length of
the order of $a_{eh}$ can be rather low: $\bar n\sim2^{-11}=1/2048$.

We performed the simulations by the quantum Monte Carlo (MC) method for 12
different densities $\bar n=2^{i-9}$ ($i=1,2...12$) in the range
$1/256\le\bar n\le8$. The number of excitons was taken to be $N=100$.
Initially, we performed the calculations by the variational MC (VMC) method
with the trial function defined by (\ref{psi}), (\ref{f}) and optimized the
parameter $\bar r_c$. At the second stage, the calculations were performed
using the ab initio diffusion MC (DMC) method \cite{b4908920}, in which, to
accelerate the convergence, the same trial function was taken but with the
parameter $\bar r_c$ being already optimized.

As is known, the inaccuracy of the trial function introduces an additional
error in the calculation of correlators by the DMC method. To reduce this
error, the data obtained by the two variants of the MC method were linearly
extrapolated, which allowed us to eliminate the first order with respect to
the small difference between the exact and trial functions of the ground state
\cite{b5203654}.

\section{Measurement results. A homogeneous system}\label{results}
In sections \ref{results} and \ref{harm}, we will use the dimensionless system
of units $\hbar=m=x_0=1$, so, $\bar n=n$, $\bar L=L$, etc.

\subsection{The the ground state energy. The chemical potential. The adiabatic
compressibility}\label{oute}
The {\it total} energy of the ground state of a two-dimensional dipolar
exciton can be represented as the sum of two terms,
\begin{equation}\label{Etot}
E_{tot}=\frac{2\pi Nne^2D}{\varepsilon}+E_0.
\end{equation}
The first term corresponds to the energy of a capacitor formed by
plane-parallel $e$ and $h$ layers spaced by the effective distance $D$. The
second term is responsible for the kinetic energy of $N$ dipoles and for their
dipolar interactions. It is equal to the energy of the ground state of the
system of these dipoles. This energy is calculated by the DMC method.

The exciton chemical potential counted from the exciton band edge can be
expressed via their total energy (\ref{Etot}) as
\begin{equation}\label{muEN}
\mu=\frac{dE_{tot}}{dN}=\mu_e+\mu_i,
\end{equation}
where $\mu_e=4\pi ne^2D/\varepsilon$ is the electrostatic contribution, and
\begin{equation}\label{muiEN}
\mu_i=\frac{dE_0}{dN}
\end{equation}
is the contribution to the chemical potential of excitons, which corresponds
to their interactions. This term defines the adiabatic compressibility of
excitons $\chi$, $1/\chi=\int(\delta\mu_i/\delta n({\bf r}))d{\bf r}$. For a
homogeneous system ($n({\bf r})\equiv n$), the variational derivative is
reduced to the ordinary derivative
($\int(\delta\mu_i/\delta n({\bf r}))d{\bf r}=d\mu_i/dn$), so
\begin{equation}\label{gmuin}
1/\chi=\frac{d\mu_i}{dn}.
\end{equation}

\begin{figure}[t]
\includegraphics[width=8.65cm,height=8.5cm]{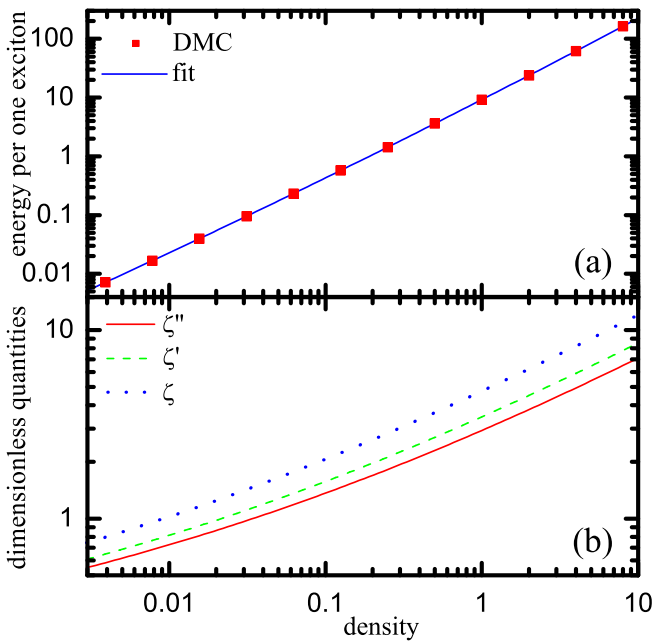}
\vskip -4mm
\caption{\small (a) Ground state energy per one exciton, $E_0/N$, as a
function of the density (squares are the DMC data) and polynomial fitting
curve (\ref{Efit}) (we set $\hbar=m=x_0=1$). (b) Dimensionless characteristics
analytically calculated from the polynomial fitting of (\ref{Efit}) as
functions of the density: energy $\zeta''$ per one exciton; contribution
$\zeta'$ to the chemical potential related to dipolar interactions; and
adiabatic compressibility $\zeta$ at $T=0$.}
\end{figure}

Figure 2(a) presents the results of the DMC simulation of the ground-state
energy of interacting dipoles per one particle, $E_0/N$, at different
densities. The calculated values lie with a high accuracy (within the limits
of 0.025\%!) on the curve of the polynomial fitting
\begin{equation}\label{Efit}
E_0/N=a_e\exp(b_e\ln n+c_e\ln^2n+d_e\ln^3n+e_e\ln^4n),
\end{equation}
where $a_e=9.218$, $b_e=1.35999$, $c_e=0.011225$, $d_e=-0.00036$, and
$e_e=-0.0000281$.

Figure 2b shows the energy per particle, $E_0/N$, the contribution to the
chemical potential $\mu_i$ that is caused by dipolar interaction, and the
adiabatic compressibility $\chi$ which were analytically calculated from the
fitting to energy (\ref{Efit}) (see Eqs.~(\ref{muiEN}), (\ref{gmuin})) and
were expressed  in terms of the quantities $\zeta''$, $\zeta'$, and $\zeta$
respectively,
\begin{equation}\label{EN}
E_0/N=\pi n\zeta'',
\end{equation}
\begin{equation}\label{mui}
\mu_i=2\pi n\zeta',
\end{equation}
\begin{equation}\label{g}
1/\chi=2\pi\zeta.
\end{equation}

We find from this simulation that, in the entire range of calculated
densities, the dimensionless adiabatic compressibility (i.e., the
dimensionless interaction) proves to be greater than or of the order of
unity,
\begin{equation}\label{zeta>1}
\zeta\ge0.774\sim1,\;\;\;\;1/256\le n\le8.
\end{equation}

In the regime of high densities ($1/4\lesssim\bar n\lesssim4$), we find that
$\zeta\gg1$ (see Fig.~2(b)). This testifies to strong correlations in the
dense gas of 2D dipolar excitons in CQWs, whose condensation was studied in
Refs.~\cite{revT,revB,T,B,M}.

At low densities, the quantity $\zeta\sim1$. This is the regime of
intermediate correlations, which was realized in Refs.~\cite{Snoke1,ss134037,%
TSQW}.

The regime of weak correlations ($\zeta\ll1$) has not been realized even at
the lowest calculated density $n=1/256$ (at which the superfluid crossover
temperature in structures considered in Refs.~\cite{revT,revB,T,B,M,Snoke1,%
ss134037,rl970103,b7400409,transp,TSQW} is not higher than 0.12 K; see Section
\ref{measur}).

\subsection{The one-body density matrix. The Bose condensate fraction. The
microscopic phonon length scale}\label{outdr}
On long-wavelength scales $r$ that exceed a certain value $r_0$ (i.e., at
$r\gg r_0$), for the polar-circle-averaged one-body density matrix
\begin{equation}\label{r1}
\rho_1(r)=\int_0^{2\pi}\langle\hat\Psi^+({\bf r})\hat\Psi(0)\rangle
d\varphi/2\pi
\end{equation}
of Bose condensed 2D dipolar excitons, the following hydrodynamic expression
holds at $T=0$ (see Appendix A):
\begin{equation}\label{r1hd}
\rho_1^{\rm hd}(r)=n_0\!\int\limits_0^{2\pi}\!\exp\!\left(\sum_{{\bf p}\ne0}
e^{i{\bf pr}}\left(\frac{\varepsilon_p}{2Np^2}-\frac1{4N}\!\right)\!-
\frac1{4N}\right)\frac{d\varphi}{2\pi},
\end{equation}
$$
r\gg r_0.
$$
In Eqs.~(\ref{r1}) and (\ref{r1hd}) $\hat\Psi({\bf r})$ is the exciton field
operator, $\langle...\rangle$ is the averaging over the exciton ground state,
$\varphi$ is the polar angle of the vector ${\bf r}$, $\varepsilon_p$ is the
excitation spectrum at $T=0$ (which we approximately determine from the
structure factor by the Feynman formula; see Sections \ref{outsp},
\ref{outep}), and $r_0$ is the characteristic microscopic phonon length scale
of the system that separates the long-wavelength (hydrodynamic, $r\gg r_0$)
and microscopic ($r\ll r_0$) ranges (Fig.~3(b)) and corresponds to the
ultraviolet limit of applicability of the hydrodynamic method. The fact that
the sum in Eq.~(\ref{r1hd}) is discrete with respect to momentum
(${\bf p}=(2\pi/L){\bf n}$, where ${\bf n}=\{n_x,n_y\}\ne0$ is the
integer-valued vector) allows us to correctly take into account both the
finiteness of the size of the system and the periodic boundary conditions.

\begin{figure}[t]
\includegraphics[width=8.65cm,height=10.0cm]{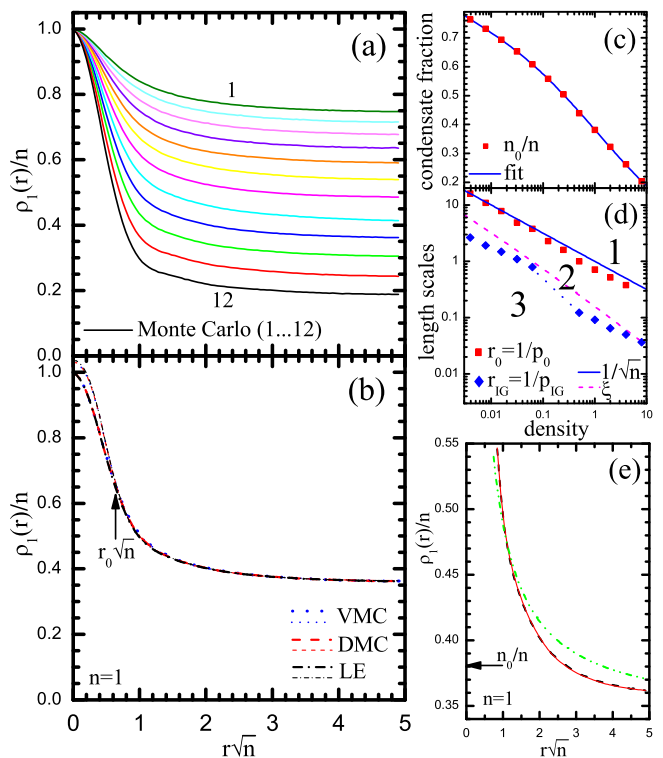}
\vskip -4mm
\caption{\small (a) One-body density matrix $\rho_1(r)$ at $T=0$ for 12
different densities ($n=2^{i-9}$, $i=1,2...12$). (b) One-body density matrix
calculated by the VMC and DMC methods, as well as linear extrapolation (LE) of
VMC and DMC results (thick curves) and hydrodynamic fits (thin curves) at
$n=1$ (the curves practically coincide). The arrow shows the characteristic
microscopic phonon length scale $r_0$ (in units of $1/\sqrt n$). (c) Bose
condensate fraction at $T=0$ ($n/n_0$) and polynomial fitting by Eq.~(19) as
functions of the density. (d) Characteristic microscopic length scales of the
system: phonon scale $r_0$, ideal-gas scale (see Section \ref{outep})
$r_{IG}$, healing length $\xi$, and interexciton distance $1/\sqrt n$ as
functions of the density. The hydrodynamic (1), intermediate (2), and
ideal-gas (3) ranges are shown. (e) One-body density matrix $\rho_1(r)$ at
$n=1$ and large $r$: DMC calculation (dashed curve), hydrodynamic calculation
(\ref{r1hd}) (solid curve), and asymptotic form (\ref{r1oo}) for the infinite
system (dot-and-dash curve). The endpoint $r/\sqrt n=5$ of the abscissa axis
corresponds to the boundary $L/2$ of the system.}
\end{figure}

The MC data for the one-body density matrix $\rho_1(r)$ (the linear
extrapolation of the VMC and DMC data) at different exciton densities are
shown in Fig.~3(a).

As shown in Fig.~3(b) the excelent coinciding of VMC and DMC results (within
the line thickness) is the evidence of a good choice of the trial wavefunction
and, consequently, a high accuracy of our simulation.

Figure 3(e) shows the one-body density matrix $\rho_1(r)$ calculated at $n=1$
and large $r$. This calculation ideally coincides (within the line thickness)
with the hydrodynamic equation (\ref{r1hd}). This testifies to the internal
consistency of our MC simulation, to the high measurement accuracy, and to the
validity of the hydrodynamic description in this range.

For comparison, Fig.~3(e) presents the asymptotic form for the one-body
density matrix at large $r$ in the corresponding infinite system
($L\to\infty$) at $n=1$,
\begin{equation}\label{r1oo}
\rho_1^{\infty}(r)=n_0^{\infty}\left(1+\frac{c_s}{4\pi n}\frac1r\right),
\;\;\;\;L\to\infty,\;r\gg r_0,1/c_s,
\end{equation}
which was obtained from Eq.~(\ref{r1hd}) by the formal replacement of the sum
over ${\bf p}$ by the integral. In Eq.~(\ref{r1oo}), the exponential is
expanded into a series, $c_s=\sqrt{n/\chi}$ is the sound velocity at $T=0$,
and $n_0^{\infty}=0.3405\pm0.001$ is the density of the Bose condensate in the
infinite system at $n=1$. The quantity $n_0^{\infty}$ was obtained by
quadratic  extrapolation of the values of $\rho_1(L/2)$ to the macroscopic
limit with respect to the powers of $1/N$ using 14 values of the total exciton
number $N$ from 25 to 200.

In Fig.~3(c), the density dependence of the the exciton Bose condensate
fraction is presented, which was obtained by the best fitting of hydrodynamic
equation (\ref{r1hd}) to the MC data. The results are described well (within
the limits of 0.005) by the following polynomial fitting curve:
\begin{equation}\label{n0fit}
n_0/n=a_0^n\exp(b_0^n\ln n+c_0^n\ln^2n+d_0^n\ln^3n),
\end{equation}
where $a_0^n=0.3822$, $b_0^n=-0.2342$, $c_0^n=-0.02852$, and
$d_0^n=-0.001594$. According to the data presented in Fig.~3(c),
$n_0/n\approx1/3$ at $n=2$, and, at $n=8$, the fraction of the Bose condensate
amounts to only $n_0/n\approx1/5$. This is indicative of strong correlations
in the dense gas of 2D dipolar excitons in CQWs in their experimentally
observed collective state. It should be specially noted that, at so small
fractions of the condensate, the application of the mean-field methods (the
Gross-Pitaevskii mean-field approach), the perturbation theory calculations,
as well as the Bogoliubov approximation, is unjustified and can be used only
for qualitative purposes.

We also note that, in the macroscopic limit ($N\to\infty$), the Bose
condensate fraction is even smaller than that at $N=100$. Thus, at $n=2$ and
$n=8$, respectively, we found that $n_0/n\approx0.29$ and $n_0/n\approx0.17$.

Figure 3(b) shows that hydrodynamic expression (\ref{r1hd}) excellently
coincides with the MC calculation of $\rho_1(r)$ at distances $r$ longer than
the characteristic microscopic phonon length scale $r_0$ (i.e., at $r>r_0$).
However, at $r<r_0$, this coincidence vanishes very rapidly. Therefore, the
crossover range of the domain of applicability of hydrodynamics for 2D dipolar
excitons turns out to be very narrow.

The behavior of the characteristic linear microscopic phonon scale $r_0$ of
the system in relation to the density is shown in Fig.~3(d). It is clearly
seen that, in the entire density range, this scale has the order of the
average interexciton distance, 
\begin{equation}\label{r0}
r_0\sim\frac1{\sqrt{n}}.
\end{equation}
In this case, the healing length $\xi=1/\sqrt{2\mu_i}$ proves to be
considerably smaller than $r_0$ at all densities (see Fig.~3(d)). This
contradicts the weak correlational behavior of 2D dipolar excitons.

\subsection{The pair distribution. The diameter of the dipole effective hard
disk. The influence of the exciton internal structure. The energy-dependent
scattering length}\label{outpd}
The polar-circle-averaged pair distribution function of excitons,
\begin{equation}\label{r2}
\rho_2(r)=\int_0^{2\pi}\langle\hat\rho({\bf r})\hat\rho(0)\rangle
d\varphi/2\pi,
\end{equation}
where $\hat\rho({\bf r})=\hat\Psi^+({\bf r})\hat\Psi({\bf r})$ is the exciton
density operator, simulated at different densities $n$ is shown in Fig.~4(a). 

\begin{figure}[t]
\includegraphics[width=8.65cm,height=8.3cm]{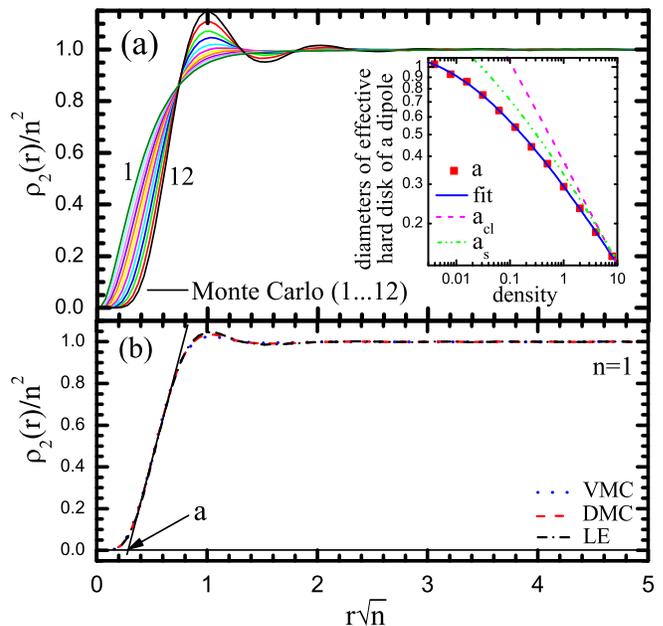}
\vskip -4mm
\caption{\small (a) Exciton pair distribution function at $T=0$ for different 
densities $n=2^{i-9}$ ($i=1,2...12$). The inset shows density dependences for
the diameter $a$ of the dipole effective hard disk, polynomial fitting curve
(\ref{afit}), energy dependent dipole scattering length, $a_s$ (\ref{as}), and
classical turning point for the dipole potential,
$a_{cl}=1/\sqrt[\mbox{\small 3}\!]{2E_0/N}$. (b) VMC and DMC results and their
linear extrapolation (LE) for pair distribution function at $n=1$ (curves
practically coincide). The arrow clarifies the definition of the diameter $a$
of the dipole effective hard disk.}
\end{figure}

At high densities ($n\ge1/8$), the pair distribution function exhibits a
clearly pronounced hump, corresponding to a short-range order. At very high
densities, there are also weaker humps. This is indicative of strong
correlations in 2D dipolar exciton system in CQWs in the low-temperature
exciton phase studied in \cite{revT,revB,T,B,M}.

The inset of Fig.~4(a) shows the density dependence of the diameter $a$ of the
effective hard disk of a dipole. We define this diameter as the distance from
the origin to the point of intersection of the short-wavelength tangent line
to the pair distribution function with the abscissa axis (see Fig.~4(b)). The
calculated points are described well (within 2\%) by the following polynomial
fitting curve:
\begin{equation}\label{afit}
a=a_a\exp(b_a\ln n+c_a\ln^2n+d_a\ln^3n),
\end{equation}
where $a_a=0.293$, $b_a=-0.324$, $c_a=-0.0123$ and $d_a=0.00096$.

The calculation of the diameter of the effective hard disk makes it possible
to evaluate the influence of the internal exciton structure, which is not
taken into account in the model of dipolar excitons. If the exciton diameter
is smaller than the diameter $a$ of the dipole effective hard disk, the
neglect of the internal structure of dipolar excitons is justified. This is
connected with the fact that the probability density of one exciton to be at a
distance of $r<a$ from another exciton, which is proportional to $\rho_2(r)$,
is small (Fig.~4(b)).

The average electron-hole distance $a_{eh}$ in the exciton can be taken to
estimate the exciton diameter in GaAs. In structure (1), which consists of
GaAs CQWs \cite{revT,T}, the average distance $a_{eh}=17$ nm; therefore, for
the dimensional exciton density $n<3.2\cdot10^{10}$ cm${}^{-2}$, $a_{eh}<a$.

If the exciton diameter is greater than $a$, the internal structure somewhat
affects the microscopic properties of dipolar excitons. However, up to
densities $n$ corresponding to the boundary between the strong and weak
coupling regimes, where the dipole barrier ceases to suppress exchange effects
(see Sections \ref{Intr} and \ref{measur}), this influence still can be
neglected.

In structure (1) \cite{revT,T}, at the maximal exciton density 
$5.17\cdot10^{10}$ cm${}^{-2}$ (see Eq.~(\ref{nmax})), corresponding to the
boundary between the strong and weak coupling regimes, we find $a=14.2$ nm
$\approx0.84a_{eh}$.

Finally, in the inset of Fig.~4(a), we present the scattering length
$a_s=a_s(p)$,
\begin{equation}\label{as}
a_s(p)=a_s^a\exp(b_s^a\ln p+c_s^a\ln^2p+d_s^a\ln^3p)
\end{equation}
of two dipoles with the relative momentum $p\sim\sqrt{2E_0/N}$ and the energy
of order of $E_0/N$. Here, $a_s^a=0.68845$, $b_s^a=-0.45897$,
$c_s^a=-0.03098$, and $d_s^a=0.002096$. Fit (\ref{as}) is based on
the calculation of Ref.~ \cite{ABKLas} and, at $0.1\le p\le20$ (i.e., at
$1/340\le n\le9.3$; see Eq.~(\ref{Efit})), coincides with it within 0.04\%. It
is seen from the inset that, at low densities, the diameter $a$ of the
effective hard disk is smaller than the scattering length $a_s$.

\subsection{The static structure factor. The sound velocity}\label{outsp}
The structure factor of 2D dipolar excitons is defined via the Fourier
transform of their pair distribution function
$\rho_2({\bf r})=\langle\hat\rho({\bf r})\hat\rho(0)\rangle$ as
\begin{equation}\label{Sbfp}
S({\bf p})=1+\int e^{-i{\bf pr}}(\rho_2({\bf r})/n-n)d{\bf r}.
\end{equation}
For a system with a finite size $L$ and with periodic boundary conditions,
this parameter is a discrete function of the momentum
(${\bf p}=(2\pi/L){\bf n}$, where ${\bf n}=\{n_x,n_y\}$ is the integer-valued
vector). However, the experiment is, as a rule, interested in large 2D Bose 
condensed exciton systems with a rather large exciton number and a complex
boundary. For their description, it is convenient to use the expression for
the structure factor in the corresponding infinite system.

For momenta $p\gg2\pi/L$, the form for the structure factor of the infinite
system, $S({\bf p})\equiv S(p)$, can approximately be obtained from
Eq.~(\ref{Sbfp}) for a flat trap of the size $L$ (with a number of excitons
equal to, e.g., $N=100$), if Eq. (\ref{Sbfp}) is smoothed by integrating over
the polar angle,
\begin{equation}\label{Sp}
S(p)=1+\int J_0(pr)(\rho_2(r)/n-n)2\pi rdr,\;\;\;\;p\gg2\pi/L.
\end{equation}
Here, it is taken into account that
$\int_0^{2\pi}e^{-ix\cos\varphi}d\varphi=2\pi J_0(x)$, where $J_0(x)$ is the
zeroth-order Bessel function. For a density of, e.g., $n=1$, the difference
between smoothed structure factors (\ref{Sp}) determined for two values of the
exciton number, $N=100$ and $N=200$, does not exceed 0.35\%. Consequently,
smoothed form (\ref{Sp}) for the structure factor approximates well this
parameter for the infinite system at $p\gg2\pi/L$.

However, at very small momenta, $p\lesssim2\pi\hbar/L$, at which effects of
the finiteness of the system manifest themselves, smoothing (\ref{Sp}) does
not yield the correct result for infinite systems. To obtain the correct
result for an infinite system at such small momenta, it is necessary to join
Eq.~(\ref{Sp}) with the long-wavelength hydrodynamic asymptotics for the
structure factor known from the general theory. For higher accuracy, we will
use the third-order long-wavelength asymptotics in momentum
\cite{Khalatn},
\begin{equation}\label{Splw}
S(p)=sp(1+\gamma p^2),\;\;\;\;p\lesssim2\pi/L,
\end{equation}
valid on hydrodynamic scales $p\ll p_0\equiv1/r_0\sim\sqrt n$ (see Section
\ref{outdr}). Here, $s=const>0$ is the parameter responsible for the
long-wavelength behavior of the structure factor, while the quantity
$\gamma=const$ determines the third order in $p$. The smooth joining (the
coincidence of the functions and of their first derivative) of the smoothed
structure factor (\ref{Sp}) and the long-wavelength asymptotics (\ref{Splw})
at intermediate momenta $2\pi/L\ll p\ll p_0$ turns out to make it possible to
rather precisely determine the value of $s$. For $N=100$, the prediction for
$s$, $s=1/\sqrt{8\pi n\zeta}$, based on the polynomial fitting (\ref{Efit}) to
the energy (see Eqs.~(\ref{muiEN})-(\ref{g}), (\ref{cs}), (\ref{csfit}))
yields the difference from the macroscopic limit that is less than by 0.08\%.
This is indicative of the adequacy of our approximations.

\begin{figure}[t]
\includegraphics[width=8.65cm,height=3.9cm]{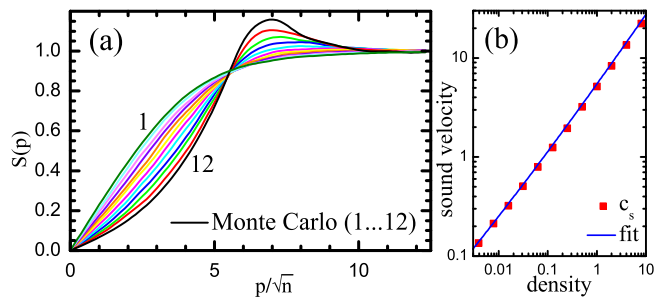}
\vskip -4mm
\caption{\small (a) Exciton structure factor at $T=0$ for different densities
$n=2^{i-9}$ ($i=1,2...12$). (b) Sound velocity determined from the slope of
the structure factor (squares) and based on fitting (\ref{Efit}) to the energy
(solid curve) as a function of the density.}
\end{figure}

Figure 5(a) depicts the structure factor of the corresponding infinite system
calculated by Eqs.~(\ref{Sp}) and (\ref{Splw}) for different densities. At
high exciton densities ($n\ge1/8$), the curves exhibit a hump corresponding
to a short-range order. At densities $1/4\lesssim\bar n\lesssim4$, this hump
is clearly seen. This testifies to strong correlations in the dense gas of 2D
dipolar excitons in CQWs studied in \cite{revT,revB,T,B,M}.

Figure 5(b) shows the dependence of the sound velocity $c_s$ in the
collisionless regime on the density, which we determine from hydrodynamic
considerations from the parameter $s$,
\begin{equation}\label{cs}
c_s=\frac1{2s}
\end{equation}
(see Eq.~(\ref{ep}) below). The calculated points for $c_s$ fall well on the
curve corresponding to the expression of this quantity via the adiabatic
compressibility,
\begin{equation}\label{csfit}
c_s=\sqrt{n/\chi}=\sqrt{2\pi n\zeta}
\end{equation}
(the dimensionless adiabatic compressibility $\zeta$ is determined by
Eqs.~(\ref{muiEN}), (\ref{gmuin}), and (\ref{g}) from fitting (\ref{Efit}) to
the energy; see Section \ref{oute}). This indicates that our DMC simulation is
internally consistent.

\subsection{The excitation spectrum. Characteristic microscopic scales of the
system}\label{outep}
In the collisionless regime, the gas phase of the Bose system is known to have
only one spectral branch. Therefore, the spectrum of elementary excitations of
Bose condensed 2D dipolar excitons in an infinite system at $T=0$ can
approximately be calculated by the Feynman formula \cite{FF},
\begin{equation}\label{ep}
\varepsilon_p=\frac{p^2}{2S(p)}.
\end{equation}
which is valid in the long-wavelength (hydrodynamic) range ($r\gg r_0$,
$p\ll p_0$), where the excitation spectrum corresponds to phonons.
Approximately, we have (see Section \ref{outsp})
$S(p)\approx sp(1+\gamma p^2)$ and $\varepsilon_p\approx c_sp(1-\gamma p^2)$.
Formula (\ref{ep}) is also true in the short-wavelength range corresponding to
an ideal gas ($r\ll r_{IG}$, $p\gg p_{IG}\equiv1/r_{IG}$), where
$S(p)\approx1$ and $\varepsilon_p\approx p^2/2$. However, on intermediate
scales ($r_{IG}\ll r\ll r_0$, $p_{IG}\gg p\gg p_0$), the Feynman formula
(\ref{ep}) is quantitatively not valid. Nevertheless, it still can be used for
qualitative estimates. The details are discussed in Appendix B.

\begin{figure}[t]
\includegraphics[width=8.65cm,height=10.0cm]{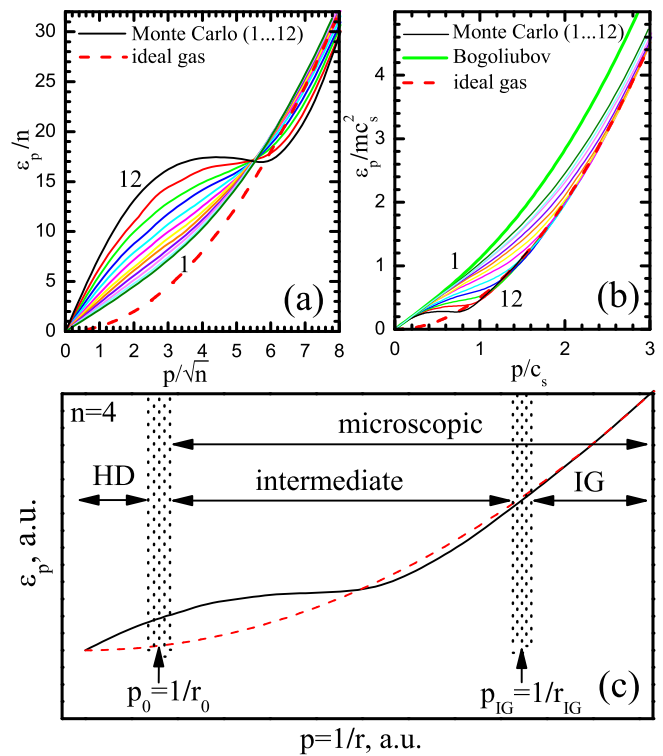}
\vskip -4mm
\caption{\small (a) Excitation spectrum at $T=0$ for different densities
$n=2^{i-9}$ ($i=1,2...12$). (b) Comparison of the excitation spectra with the
Bogolyubov shape. (c) Excitation spectrum for the density $n=4$ (solid curve)
and the spectrum of the ideal gas (dashed curve). The hydrodynamic (HD;
$r\gg r_0$, $p\ll p_0$), ideal-gas (IG; $r\ll r_{IG}$, $p\gg p_{IG}$),
intermediate ($r_0\gg r\gg r_{IG}$, $p_0\ll p\ll p_{IG}$), and microscopic
($r\ll r_0$, $p\gg p_0$) ranges are shown.}
\end{figure}

In Figs.~6(a) and 6(b), the excitation spectra in an infinite gas of 2D
dipolar excitons analytically calculated by Eq.~(\ref{ep}) for different
densities are shown. It is seen that, in the entire density range used in the
calculations, the excitation spectrum is far from the Bogolyubov shape
$\varepsilon_p^B=\sqrt{p^4/4+c_s^2p^2}$ (although, at low densities, it
qualitatively resembles the Bogolyubov spectrum). At high densities
($1/4\lesssim\bar n\lesssim4$), effects of strong correlations are clearly
seen. At the maximal density used in the calculation, $n=8$, a roton minimum
appears in the spectrum.

The ranges of long-wavelength (hydrodynamic), short-wavelength (ideal-gas),
intermediate, and microscopic scales are clearly seen in Fig.~6(c). The
diagrams of the long-wavelength, intermediate, and microscopic ranges were
presented above in Fig.~3(d) (see Section \ref{outdr}).

\subsection{Temperature dependence of the local superfluid density.
Quasicondensation and BKT transition temperatures}\label{outnl}
In the quasicondensed \cite{Popov} phase of excitons far from the crossover
between the quasi-classical regime and the quasicondensed phase, elementary
excitations form a nearly ideal gas. In this case, the temperature dependence
of the fraction of the {\it local} (vortex-unrenormalized \cite{VR})
superfluid component of quasicondensed 2D dipolar excitons can approximately
be calculated by the Landau formula \cite{AGD},
\begin{equation}\label{nl}
\frac{n_l(T)}n=1+\int\frac{p^2}2\frac{dn(\varepsilon_p^T)}
{d\varepsilon_p^T}\frac{pdp}{2\pi n}g_{ex},\;\;\;\;T_q-T\sim T_q.
\end{equation}
Here, $n_l$ is the local superfluid exciton density (in the $g_{ex}$ spin
degrees),
\begin{equation}\label{nep}
n(\varepsilon_p^T)=\frac1{e^{\varepsilon_p^T/T}-1}
\end{equation}
is the Bose distribution of the ideal gas of elementary excitations with the
spectrum $\varepsilon_p^T$ at the exciton temperature $T$, and $T_q$ is the
estimate for the temperature of quasicondensation crossover, at which the
local superfluid component and the local long-range order [57] vanish
gradually.

The {\it global} superfluid component (which takes into account the vortex
renormalization \cite{jc061181,VR}) was calculated in \cite{LKW}. To calculate
the local component, one can set
$\varepsilon_p^T\approx\varepsilon_p^{T=0}=\varepsilon_p$ in Eqs.~(\ref{nl})
and (\ref{nep}). The details of the calculation of the local superfluid
component are discussed in Appendix B.

\begin{figure}[t]
\includegraphics[width=8.65cm,height=11.1cm]{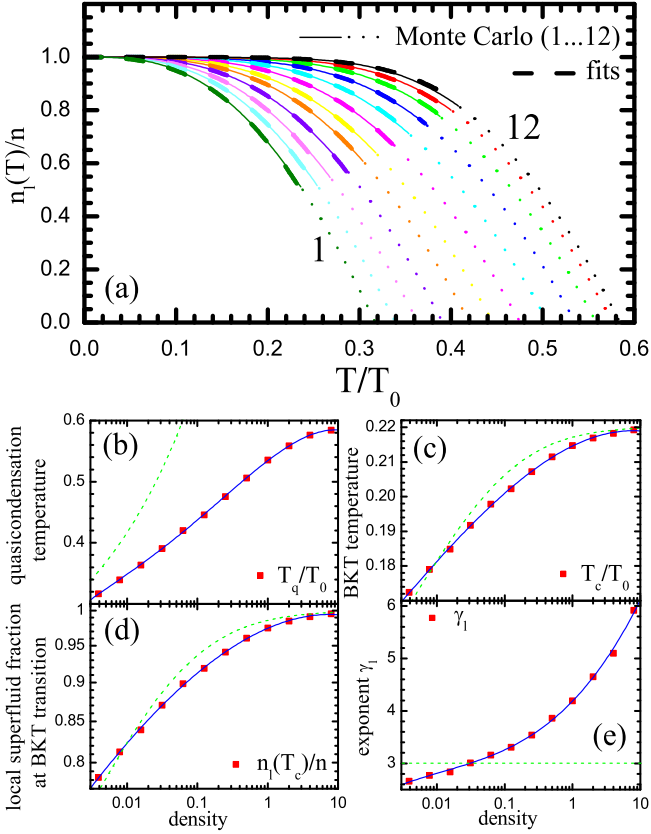}
\vskip -4mm
\caption{\small (a) Temperature dependence of the local superfluid exciton
fraction in the $g_{ex}=4$ spin degrees for different densities $n=2^{i-9}$
($i=1,2...12$) (solid curves), low-temperature power-law fitting curves
determined by Eq.~(\ref{nlfit}) (dashed curves); and extrapolation to the
range of the quasicondensed crossover (dotted curves). (b-d) Density
dependences (squares) for (b) the quasicondensation temperature $T_q$, (c) the
BKT transition temperature $T_c$, and (d) the local superfluid fraction at
$T=T_c$; solid and dashed curves correspond to the polynomial fitting and to
phonon-model calculation, respectively. (e) Exponent $\gamma_l$ in
Eq.~(\ref{nlfit}) as a function of the density: (squares) the fit for each
density, (solid curve) the general fitting by Eq.~(\ref{glfit}), and (dashed
curve) the calculation within the phonon model.}
\end{figure}

The dependence $n_l(T)/n$ analytically calculated by means of
Eqs.~(\ref{ep})-(\ref{nep}) with $\varepsilon_p^T=\varepsilon_p$ at different
densities is shown in Fig.~7(a), where the quantity
\begin{equation}\label{T0}
T_0=g_{ex}T_{deg}=2\pi n\;\;\;\;(g_{ex}=4\mbox{ in GaAs}).
\end{equation}
denotes the degeneracy temperature for the corresponding spin-polarized
excitons.

In the temperature range $0<T<1.2T_c$, with an error less than 0.003, this
calculation of the local superfluid fraction coincides with the following
power-law fitting curve:
\begin{equation}\label{nlfit}
n_l(T)/n=1-(1-n_l(T_c)/n)(T/T_c)^{\gamma_l},
\end{equation}
\begin{equation}\label{glfit}
\gamma_l=a_l^{\gamma}+b_l^{\gamma}\ln n+c_l^{\gamma}\ln^2n+d_l^{\gamma}\ln^3n,
\end{equation}
where $a_l^{\gamma}=4.186$, $b_l^{\gamma}=0.5825$, $c_l^{\gamma}=0.0941$, and
$d_l^{\gamma}=0.00706$, while $T_c<T_q$ is the temperature of the BKT
superfluid transition \cite{Berezinskii,jc061181,jc071046}, at which the
global superfluid component in the infinite system vanishes jump-wise
\cite{jc061181,rl391201}. For $T_c$, we use the equation \cite{jc061181,%
rl391201} (see also Appendix B)
\begin{equation}\label{Tc}
T_c=\frac{\pi n_l(T_c)}{2\tilde\epsilon},
\end{equation}
in which the dielectric permittivity $\tilde\epsilon$ of vortex pairs at the
BKT transition can be taken from the 2D $x$-$y$ model
($\tilde\epsilon=1.1349\pm0.0005$; see \cite{LKW}).

In Figs.~7(b)-(d), we present the analytically calculated points (denoted by
squares) for the temperatures of quasicondensation, $T_q$ (see Appendix B),
and of the superfluid BKT transition, $T_c$, as well as for the local
superfluid fraction, $n_l(T_c)/n$, at the BKT transition in the infinite
system. These points fall very well on the corresponding polynomial fitting
curves (within the limits 0.0025, 0.0005, and 0.0025, respectively),
\begin{equation}\label{Tqfit}
T_q/T_0=a_q^T+b_q^T\ln n+c_q^T\ln^2n+d_q^T\ln^3n+e_q^T\ln^4n,
\end{equation}
\begin{equation}\label{Tcfit}
T_c/T_0=a_c^T+b_c^T\ln n+c_c^T\ln^2n+d_c^T\ln^3n,
\end{equation}
\begin{equation}\label{nlcfit}
n_l(T_c)/n=a_c^l+b_c^l\ln n+c_c^l\ln^2n+d_c^l\ln^3n.
\end{equation}
Here, $a_q^T=0.5344$, $b_q^T=0.038$, $c_q^T=-0.00395$, $d_q^T=-0.001146$, and
$e_q^T=-0.000086$; $a_c^T=0.2146$, $b_c^T=0.00414$, $c_c^T=-0.000877$, and
$d_c^T=-0.0000446$; $a_c^l=0.9745$, $b_c^l=0.0188$, $c_c^l=-0.00398$, and
$d_c^l=-0.000202$.

According to our results, the phonon model, which is widespread in the
literature \cite{SFB,LB} (in which $\varepsilon_p\equiv c_sp$), describes
rather well the temperature of the BKT transition (Fig.~7(c)) but fails to
adequately describe the temperature of the quasicondensation crossover
(Fig.~7(b)). (The phonon model does not take into account the roton bend of
the spectra on intermediate scales (see Figs.~6(a),(c)), which makes the main
contribution at high densities and $T\sim T_q$.)

It is important that, at high densities ($1/4\lesssim\bar n\lesssim4$), the
local superfluid fraction at the BKT transition, $n_l(T_c)/n$, is close to
unity. This is inductive of strong correlations in the dense 2D dipolar
exciton gas in CQWs \cite{revT,revB,T,B,M} (see also Ref. \cite{nl=n}).

In addition, for the spin-depolarized dense 2D dipolar exciton gas in GaAs
($g_{ex}=4$), the temperature of quasicondensation $T_q\sim0.5T_0$ (see
Fig.~7(b)) is approximately two times higher than the temperature of
degeneracy $T_{deg}=T_0/g_{ex}=0.25T_0$ (see Eq.~(\ref{T0})), whereas the BKT
transition temperature $T_c\approx0.2T_0$ (see Fig.~7(c)) is only slightly
lower than $T_{deg}$. This strongly differs from models widespread in the
literature of ideal and weakly correlated 2D exciton gases in QWs for which
these temperatures are logarithmically small compared to the degeneracy
temperature $T_{deg}$ \cite{b3704936}.

Therefore, the condensed state of 2D dipolar excitons in CQWs and SQWs {\it
can be experimentally obtained much simpler} than in the case of weakly
correlated excitons.

\section{A harmonic trap at $T=0$}\label{harm}
Modern experimental methods are capable of harmonic trapping of 2D dipolar
excitons in the plane of a quantum well. This trapping can be obtained with
the help of an inhomogeneous compression of the sample, caused by the pressure
of a tip on its surface \cite{ss134037,rl970103}, as well as with the help of
an inhomogeneous electric field in electrostatic traps \cite{revT,b7400409,%
jl800185,mj360940,a0990604,prb076085304}.

In the latter experiments on Bose condensation of excitons, their number in a
harmonic trap is large, $N\gg1$ \cite{revT,ss134037,rl970103,jl800185}. In
this case, the phonon microscopic length scale $r_0\sim1/\sqrt{n(0)}$ (see
Eq.~(\ref{r0}); $n({\bf R})$ is the total exciton profile in the trap) proves
to be significantly smaller than the trap size $L_{TF}\sim\sqrt{N/n(0)}$
($L_{TF}=\sqrt{2N/\pi n(0)}$ is the Thomas-Fermi (TF) radius). In the
calculated density range, other microscopic scales ($\xi$, $r_{IG}$, $a$,
etc.) are even smaller than $r_0$ (see Section \ref{results}). Therefore,
microscopic properties of excitons in a trap can be formed on local scales,
along which the potential of the trap varies continuously.

Moreover, if the exciton number $N$ in a trap is substantially greater than
$4\pi^2\approx40$ the hydrodynamic range $p\ll1/r_0\sim\sqrt{n(0)}$
considerably exceeds the momentum discreteness step
$2\pi/L_{TF}\sim2\pi\sqrt{n(0)/N}\ll\sqrt{n(0)}$, which is determined by
boundary effects in the homogeneous system with the size $L=L_{TF}$ (see
Sections \ref{outdr} and \ref{outsp}). Therefore, in the exciton system in a
large trap containing $N\gg40$ excitons, the sound range of momenta,
$2\pi/L_{TF}\ll p\ll1/r_0$ does exist and is not masked by finite-size
effects, arising on scales $p\lesssim2\pi/L_{TF}$.

Therefore, a sufficiently large harmonically trapped 2D dipolar exciton system
at $T=0$ can be fairly well described in the local density approximation
(LDA) \cite{LDA}. In this approximation, the local microscopic exciton
properties near the point ${\bf R}$ of the trap are approximately replaced by
the properties of the corresponding homogeneous system with the density $n$
equal to the exciton density $n({\bf R})$ in the trap at the point ${\bf R}$.

Thus, within the framework of the LDA, the total exciton density profile in a
(symmetric) trap $n(R)$ is determined from the TF equation, which, at $T=0$,
takes the form
\begin{equation}\label{TF}
V(R)+\mu(n(R))=\mu,
\end{equation}
where $V(R)=(m/2)\omega_{ho}^2R^2$ is the potential of the trap with the
oscillator frequency $\omega_{ho}$, $\mu$ is the chemical potential of
excitons in the trap counted from the exciton band edge, and $\mu(n(R))$ is
the local chemical potential of excitons in the trap equal to their chemical
potential in the homogeneous system with the density $n=n(R)$.

The density of exciton Bose condensate at $T=0$ in a harmonic trap can be
calculated in the LDA as
\begin{equation}\label{n0R}
n_0(R)=n_0(n(R)),
\end{equation}
where $n_0(n(R))$ is the Bose condensate density in the homogeneous system
with the total density $n=n(R)$.

\begin{figure}[t]
\includegraphics[width=8.65cm,height=8.5cm]{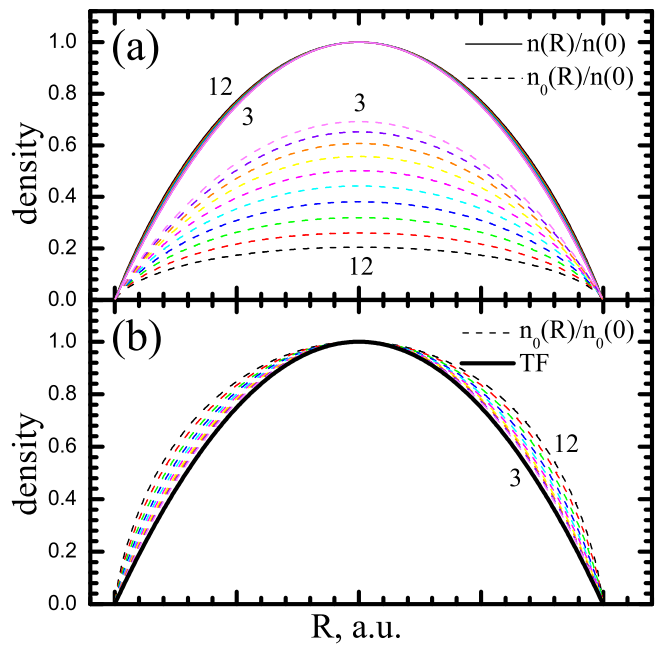}
\vskip -4mm
\caption{\small (a) Total ($n(R)/n(0)$) and Bose condensate ($n_0(r)/n(0)$)
exciton profiles in a harmonic trap in the LDA at $T=0$ and
$\zeta_e'\equiv2e^2D/\varepsilon=10$ for different densities at the center
($n(0)=2^{i-9}$, $i=3,4... 12$). (b) Comparison of the Bose condensate
profiles at different $n(0)$ with the Thomas-Fermi inverted parabola.}
\end{figure}

Figure 8(a) shows the total and Bose condensate exciton profiles in a harmonic
trap at $T=0$ analytically calculated in the LDA for different densities
$n(0)$ at the trap center. The quantities $\mu(n)$ and $n_0(n)$ in the
infinite homogeneous system were approximately calculated using the
corresponding fits (\ref{Efit}) and (\ref{n0fit}) (see also (\ref{muEN}) and
(\ref{muiEN})) for the homogeneous finite system consisting of $N=100$
excitons.

It is clearly seen from this figure that, at
$\zeta_e'\equiv\mu_e/T_0\equiv2e^2D/\varepsilon=10$, the total density $n(R)$
for all $n(0)$ nearly ideally fall on the profile of the inverted TF parabola
(see Fig.~8(a)),
\begin{equation}\label{nR}
n(R)\approx n(0)(1-R^2/L_{TF}^2)\theta(L_{TF}-R)
\end{equation}
where $\theta(x)=0$ at $x<0$ and $\theta(x)=1$ at $x>0$. However, this fact is
determined not so much by a weak nonlinearity of the collisional contribution
to the chemical potential as by a large value of the electrostatic
contribution $\mu_e=\zeta_e'T_0=10T_0$, which is linear in the density (see
Eq.~(\ref{T0})). (The values of the electrostatic contribution to the chemical
potential $\zeta_e'\approx10$ were realized in experiments
\cite{revT,revB,T,B,M,rl970103,b7400409,transp}.)

However, the shape of the Bose condensate profile $n_0(R)$ at high densities
($1/4\lesssim\bar n\lesssim4$) appreciably differs from that of the inverted
parabola $n_0(0)(1-R^2/L_{TF}^2)\theta(L_{TF}-R)$, Fig.~8(b). This is evidence
in favor of strong correlations in the dense 2D dipolar exciton gas.

Note that the LDA makes it possible to predict the frequencies of collective
oscillations of the compression mode, when the excitation is caused by a sharp
change in the frequency of the trapping potential \cite{a7200620}. The
frequency of this mode depends on the particular form of the two-particle
interaction (in contrast to the mode that is caused by the displacement of the
center of mass and that depends only on the trap frequency). This makes it
possible to experimentally investigate the equation of state (see
Eq.~(\ref{Efit})). Another important quantity measured in experiment is the
release energy. A rapid switching-off of a trap turns the trapping potential
to zero but practically has no effect on the kinetic energy and the part of
the potential energy related to the pair interaction. The released energy can
be measured during the spread. Using the LDA, it is possible to estimate the
release energy \cite{a7200620}.

\section{The possibility of the experimental observation of strong exciton
correlations}\label{lum}
Effects of strong correlations in a dense 2D dipolar exciton gass in CQWs can
be found upon observation of the particular features of exciton luminescence
in a longitudinal magnetic field.

Indeed, in the absence of a magnetic field, according to the momentum
conservation for one-photon exciton recombination, the momentum $p$ of a
recombining exciton is equal to the projection of the momentum of an emitted
photon onto the QW plane, $(\hbar\omega/c_0)\sin\theta$. Here, $\theta$ is the
angle between the emitted photon and a normal in a vacuum, $c_0$ is the light
velocity in a vacuum, and $\omega$ is the photon frequency (now we pass to the
ordinary, dimensional, units). However, if a longitudinal magnetic field
$H_{\parallel}$ is applied to the system, the dispersion curve $\varepsilon_p$
of 2D dipolar excitons is {\it shifted} by the quantity
$p_H=eDH_{\parallel}/c_0$ \cite{BH} (this effect is equivalent to the
excitation of diamagnetic currents in a system of coupling $e$ and $h$ by a
longitudinal magnetic field \cite{b6502304} that was described in \cite{LY}).
As a result, the relation between the emitted-photon angle $\theta$ and the
recombined-exciton momentum $p$ takes the form
\begin{equation}\label{ppH}
(\hbar\omega/c_0)\sin\theta=|{\bf p}-{\bf p}_H|.
\end{equation}
If we are interested in luminescence along the normal ($\theta=0$), then,
$p=p_H$ (see Eq.~(\ref{ppH})), so, $\varepsilon_p=\varepsilon_{p_H}$. Hence,
for the spectral-angular luminescence along the normal in the field
$H_{\parallel}$ at $T=0$ we obtain
\begin{equation}\label{It0o}
I_{\theta=0}^H(\omega)=I_{\theta=0}^H
\delta(\omega-\Omega+\varepsilon_{p_H}/\hbar),\;\;\;\;p_H\gg p_T,
\end{equation}
where $I_{\theta=0}^H$ is the spectrally integrated angular luminescence along
the normal in the field $H_{\parallel}$, $\Omega$ is the exciton resonance
frequency (for the luminescence at $H_{\parallel}=0$ and $\theta=0$), and
$p_T=T/c_s$ is the characteristic thermal momentum of Bose condensed excitons,
which separates the ranges of thermal $p\ll p_T$ and zero-temperature
$p_T\ll p\ll p_0$ quasicondensate phase fluctuations \cite{Popov}.

Eq.~(\ref{It0o})) shows that the measurement of the luminescence line
redshift \cite{Butovblueshift} $\varepsilon_{p_H}/\hbar$ in the magnetic field
$H_{\parallel}=c_0p_H/eD$ makes it possible {\it to directly determine the
excitations spectrum} $\varepsilon_p$. At $D=12.3$ nm, $m=0.22m_0$,
$\varepsilon=12.5$ \cite{revB,B}, $n=2\cdot10^{10}$ cm${}^{-2}$, and
$H_{\parallel}=8$ T, we have
$$
p_H/\hbar=1.49\cdot10^6\mbox{ cm}{}^{-1}=10.5\sqrt n\approx p_{IG}/\hbar
$$
(see Section \ref{measur} and Fig.~3(d)). Therefore, in structure (2)
\cite{revB,B} containing an exciton gas with the density $n=2\cdot10^{10}$
cm${}^{-2}$, the field $H_{\parallel}=8$ T covers both the sound
(hydrodynamic, $p\ll p_0$) and the intermediate ($p_0\ll p\ll p_{IG}$)
spectral ranges (see Fig.~6(c)). In this case, at $T=0.2T_0=1$ K, $m=0.22m_0$,
and $\zeta=3.7$ (see Fig.~2(b) and Sections \ref{measur}, \ref{oute}), the
thermal momentum $p_T$ corresponds to the field
$$
H_{\parallel}^T=\frac{c_0T}{eD}\sqrt{\frac m{\zeta T_0}}\approx0.2\mbox{ T}.
$$

Knowing the slope of the measured spectrum $\varepsilon_{p_H}$, one can
calculate the sound velocity $c_s\approx\varepsilon_{p_H}/p_H$ ($p_H\ll p_0$;
see Fig.~6(a)). This yields the dimensionless adiabatic compressibility (see
(\ref{csfit}))
$$
\zeta=m^2c_s^2/2\pi\hbar^2n.
$$

Alternatively, the adiabatic compressibility $\zeta$ can be found from the
measurement of the spectrally integrated luminescence along the normal in the
magnetic field, $I_{\theta=0}^H$.

Indeed, the spectrally integrated luminescence into the solid angle
$d\Omega=\sin\theta d\varphi d\theta$ near the normal is connected with the
momentum distribution of excitons
$n_{p_H}\equiv\int e^{-\frac i{\hbar}{\bf p}_H{\bf r}}\rho_1({\bf r})d{\bf r}$
(i.e., the number of excitons with the momentum $p_H$) by the following
relation:
\begin{equation}\label{It}
I_{\theta=0}^Hd\Omega=\kappa\frac{\hbar\omega}{\tau_0}n_{p_H}
\frac{d{\bf p}_H}{(2\pi\hbar)^2}.
\end{equation}
Here, $\tau_0$ is the lifetime of an isolated spin-depolarized exciton in the
ground state and $\kappa=1/2$ ($\kappa=1$) in the case of the measurement of
the luminescence only on one side (on both sides) of the QW plane.

On scales $p_T\ll p_H\ll p_0$ that are the most interesting to us, the
momentum distribution at $T=0$ is given by the Fourier transform of
Eq.~(\ref{r1oo}),
\begin{equation}\label{np}
n_{p_H}=\frac{n_0}n\frac{mc_s}{2p_H},\;\;\;\;p_T\ll p_H\ll p_0.
\end{equation}
From Eqs.~(\ref{It}) and (\ref{np}), we find the sought relation between the
angular luminescence along the normal, $I_{\theta=0}^H$, and the dimensionless
adiabatic compressibility $\zeta$,
\begin{equation}\label{IpHz}
I_{\theta=0}^H=I_0\frac{q_r^2c_0}{4\pi eD\sqrt{mT_0}}
\frac{\sqrt{\zeta}}{H_{\parallel}},\;\;\;\;
p_T\ll eDH_{\parallel}/c_0\ll\hbar\sqrt n.
\end{equation}
Here, $I_0=\kappa(\hbar\Omega/\tau_0)n_0$ is the exciton Bose condensate
luminescence at $T=0$. At $n=2\cdot10^{10}$ cm${}^{-2}$, $D=12.3$ nm,
$T=0.1T_0=0.5$ K, $m=0.22m_0$, $\varepsilon=12.5$, and $\zeta=3.7$, the
momentum range $p_T\ll eDH_{\parallel}/c_0\ll p_0$ in Eq.~(\ref{IpHz})
corresponds to the magnetic-field range
$$
\frac{c_0T}{eD}\sqrt{\frac m{\zeta T_0}}\ll H_{\parallel}\ll
\frac{c_0\hbar\sqrt n}{eD},
$$
or 0.1 T $\ll H_{\parallel}\ll$ 0.75 T.

The luminescence intensity of the Bose condensate at $T=0$ appearing in
Eq.~(\ref{IpHz}) can be approximately calculated as the contribution of
thermal phase fluctuations to the quasicondensate luminescence at low
temperatures (see Eq.~(\ref{It})),
\begin{equation}\label{I0}
I_0\approx\int_0^{p_T}I_{\theta=0}^H\frac{2\pi p_Hdp_H}{q_r^2}=
\kappa\frac{\hbar\Omega}{\tau_0}n_0,\;\;\;\;T\ll T_q,
\end{equation}
where $q_r=\hbar\Omega/c_0$ is the radiation bandwidth in a vacuum. Here, it
is taken into account that the exciton gap $\hbar\Omega\sim1$ eV considerably
exceeds all characteristic energy scales for excitons
$\varepsilon_{p_H},T\lesssim1$ meV $\ll\hbar\Omega$, so,
$\hbar\omega\approx\hbar\Omega$. It is also taken into account that
$dp_H=q_rd\sin\theta=q_rd\theta$ at $p_H=p$ and $\theta=0$ (see
Eq.~(\ref{ppH})). (Eq.~(\ref{I0}) can be obtained by extending the method
described in Appendix A to the case $T\ne0$, see Refs.~\cite{PFCT,Popov}.)

Finally, by measuring the quantity (\ref{I0}), one can find the important
parameter of exciton correlations, namely, the Bose condensate density at
$T=0$,
$$
n_0=\frac1{\kappa}\frac{\tau_0}{\hbar\Omega}I_0.
$$

\section{Conclusion}\label{concl}
By means of the ab initio simulation and analytical calculations, we studied
in detail the microscopic properties of Bose condensed superfluid 2D dipolar
excitons in QWs. For a homogeneous exciton system, we numerically calculated
the ground-state energy, the one-body density matrix, and the pair
distribution function. Based on these numerical calculations for the
homogeneous system, we analytically found the structure factor, the excitation
spectrum, the Bose condensate density, the temperature dependence of the local
superfluid density, the chemical potential, the adiabatic compressibility, the
sound velocity, and the characteristic microscopic length scales (the phonon
scale, the healing length, the scattering length, the diameter of the
effective hard disk of a dipole, etc.), as well as the quasicondensation
crossover and Berezinskii-Kosterlitz-Thouless transition temperatures. For a
harmonic trap with a large excton number, we analytically calculated the total
and Bose condensate profiles.

We showed that, in all experiments on Bose condensation performed at present,
2D dipolar excitons in coupled QWs prove to be strongly correlated. As a
consequence, exciton Bose condensation and superfluidity can experimentally be
achieved much easily than this is predicted by the models of an ideal and
weakly correlated exciton gases.

The results obtained for the ground-state energy $E_0/N$, chemical potential
$\mu=\mu_i+\mu_e$, the adiabatic compressibility $\chi$, the sound velocity
$c_s$, the excitation spectrum $\varepsilon_p$, and the Bose condensate
$n_0$ and local superfluid $n_l(T)$ densities, as well as for the
characteristic microscopic phonon length scale $r_0$, can be used as reference
data in construction of quantitative hydrodynamic theory of the optical
properties of 2D dipolar excitons in QWs, in particular, of their
luminescence, coherence, and nonlinear effects.

\appendix
\section{Long-wavelength $ $ hydrodynamic asymptotics for the one-body
density matrix}
On long-wavelength scales ($r\gg r_0$), the one-body density matrix
(\ref{r1}) of Bose condensed 2D dipolar excitons at $T=0$ can be calculated
within the framework of the hydrodynamic method in quantum field theory
\cite{PFCT},
\begin{equation}\label{r1cum}
\rho_1({\bf r})=C
e^{-\langle(\hat\varphi({\bf r})-\hat\varphi(0))^2/2\rangle}=
n_0e^{\langle\hat\varphi({\bf r})\hat\varphi(0)\rangle},\;\;\;\;r\gg r_0.
\end{equation}
Here, $\hat\varphi({\bf r})$ is the phase of the exciton field operator
$\hat\Psi({\bf r})$
($\hat\Psi({\bf r})=e^{i\hat\varphi({\bf r})}\sqrt{\hat\rho({\bf r})}$), which
satisfies the commutation relation
$[\hat\varphi({\bf r}),\hat\varphi({\bf r}')]=0$, and
$C=n_0e^{\langle(\hat\varphi(0))^2\rangle}=const$ does not depend on
${\bf r}$. We assume that an ultraviolet cutoff is imposed (in this case, the
quantities $C$ and $\langle(\hat\varphi(0))^2\rangle$ diverge in the
ultraviolet limit, but, at $r\gg r_0$, $n_0$ and
$\langle\hat\varphi({\bf r})\hat\varphi(0)\rangle$ are finite in the
large-momentum limit).

At $T=0$, the phase-phase correlator
$\langle\hat\varphi({\bf r})\hat\varphi(0)\rangle$ in Eq.~(\ref{r1cum}) can be
expressed via the phase-phase Green's function (in dimensional units),
$$
D_p^{\varphi\varphi}(\omega)=-i\int\langle\hat{\rm T}[
\hat\varphi({\bf r},t)\hat\varphi(0,0)]\rangle
e^{-\frac i{\hbar}{\bf pr}+i\omega t}d{\bf r}dt,
$$
as follows:
\begin{equation}\label{ff0}
\langle\hat\varphi({\bf r})\hat\varphi(0)\rangle=
\frac1{L^2}\sum\nolimits_{{\bf p}\ne0}e^{\frac i{\hbar}{\bf pr}}
\int iD_p^{\varphi\varphi}(\omega)\frac{d\omega}{2\pi}.
\end{equation}
where $\hat{\rm T}$ is the chronological operator.

In the hydrodynamic (long-wavelength) approximation, in which the hydrodynamic
Hamiltonian acquires the form local in time, the function
$D_p^{\varphi\varphi}(\omega)$ is given by \cite{Popov,PHD}
\begin{equation}\label{Dffp0}
D_p^{\varphi\varphi}(\omega)=
\frac{\hbar m\varepsilon_p^2/n_sp^2}{\hbar^2\omega^2-\varepsilon_p^2+i\delta},
\;\;\;\;p\ll p_0,
\end{equation}
where $n_s$ is the global superfluid exciton density (in the $g_{ex}$ spin
degrees).

In the calculated density range ($nx_0^2\le8$), excitons are in the gas phase
\cite{rl980605}. In addition, we neglect the external random potential, which
is unavoidably present in heterostructures. In this case, the global
superfluid exciton density at $T=0$ coincides with the total density (see
\cite{jc061181} and Section \ref{outnl}),
\begin{equation}\label{ns=n}
n_s=n.
\end{equation}

From Eqs.~(\ref{ff0})-(\ref{ns=n}) we obtain the following equation for the
correlator $\langle\hat\varphi({\bf r})\hat\varphi(0)\rangle$ at $T=0$
$$
\langle\hat\varphi({\bf r})\hat\varphi(0)\rangle=
\frac1{L^2}\sum\nolimits_{{\bf p}\ne0}
e^{\frac i{\hbar}{\bf pr}}\frac{m\varepsilon_p}{2np^2}=
$$
$$
=\frac1{L^2}\sum_{{\bf p}\ne0}e^{\frac i{\hbar}{\bf pr}}
\frac{m\varepsilon_p}{2np^2}-
\frac1{4N}\sum_{\bf p}e^{\frac i{\hbar}{\bf pr}}=
$$
\begin{equation}\label{fft}
=\sum_{{\bf p}\ne0}e^{\frac i{\hbar}{\bf pr}}
\left(\frac{m\varepsilon_p}{2Np^2}-\frac1{4N}\right)-\frac1{4N},
\end{equation}
$$
r\gg r_0.
$$
Here, we used the fact that, at $r\gg r_0$ (i.e., at ${\bf r}\ne0$), the
$\delta$-like contribution
$$
\frac1{4N}\sum_{\bf p}e^{\frac i{\hbar}{\bf pr}}=
\frac1{4n}\sum_{\bf n}\delta({\bf r}-{\bf n}L)
$$
is zero in a box of a size $L$ (${\bf n}=\{n_x,n_y\}$ is the integer-valued
vector).

By substituting (\ref{fft}) into (\ref{r1cum}) and averaging over the polar
angle of the vector ${\bf r}$, we arrive at the sought hydrodynamic equation
(\ref{r1hd}) for the one-body density matrix of Bose condensed 2D dipolar
excitons at $T=0$, which, on long-wavelength scales ($r\gg r_0$), coincides
excellently with our DMC simulation (see Fig.~3(e)).

In conclusion, we note that, in interacting systems with the finite scattering
length, the spectrum of excitations $\varepsilon_p$, at $p\to\infty$,
exponentially approaches the spectrum of the ideal gas
$\varepsilon_p=p^2/2m+O(e^{-p/p_1})$ (where $p_1=const>0$; see also item (2)
in Appendix B). Consequently, at $p\to\infty$, the quantity
$m\varepsilon_p/2Np^2-1/4N$ under the summation sign in (\ref{fft}) is of the
order of $O(e^{-p/p_1})$. Therefore, Eq.~(\ref{fft}) for the phase-phase
correlator converges in the ultraviolet limit. This means that the well-known
ultraviolet cutoff \cite{Popov} in quantum-field hydrodynamics at the boundary
of the long-wavelength band ($r\sim r_0$, $p\sim p_0$) proves to be redundant
in (\ref{fft}): an infinitely small cutoff suffices to ensure convergence.

However, one should bear in mind that the hydrodynamic (long-wavelength)
approximation fails to adequately describe the phase-phase correlator on
microscopic scales $r\ll r_0$, $p\gg p_0$. So, the range $p\gg p_0$ is
incorrectly taken into account in (\ref{fft}). As a result, the hydrodynamic
approach used in this study introduces a certain error in the calculation of
the one-body density matrix and, therefore, in the Bose condensate density.
However, direct calculation shows that, if a continuous finite ultraviolet
cutoff is introduced in (\ref{fft}) at $p\sim p_0$, the correction to the Bose
condensate density will rapidly vanish at large $N$. The numerical calculation
of the sum in (\ref{fft}) shows that, at $N=100$, this correction is of the
order of 0.001.

\section{Applicability conditions and calculation details}
Here, we will discuss subtle aspects of calculations of the excitation
spectrum, the local superfluid density, and the quasicondensed and superfluid
transition temperatures.

(1) As Boronat et al. showed \cite{FF}, in strongly correlated systems on
intermediate scales, the true spectrum lies lower than the spectrum calculated
by the Feynman formula (although, the weaker the correlations, the more exact
the Feynman formula). Therefore, in the strong-correlation regime, our
calculation of the local superfluid fraction, as well as the quasicondensation
$T_q$ and the BKT transition $T_c$ temperatures is valid only qualitatively.

(2) For simplicity, the excitation spectrum was estimated omitting the effects
connected with a large damping of elementary excitations above the endpoint of
the spectrum \cite{AGD}.

(3) The Landau formula is valid only in the collisionless (not hydrodynamic)
regime when elementary excitations weakly interact, and their damping is
small. This is definitely not the case above the spectrum endpoint $p_c$. In a
dense system, the spectrum endpoint lies high (see Fig.~6(a) and \cite{AGD}),
so that the range $p>p_c$ is cut in (\ref{nl}) by the Bose exponential
(\ref{nep}). At low densities, the spectrum endpoint $p_c$ corresponds to
smaller frequencies, but a significant damping appears only above the sound
range of the spectrum. This range is also cut by the Bose exponential (see
(\ref{nl})-(\ref{T0}) and Figs.~6(a), 7(b)). However, at very low densities
$\bar n$, where excitons are weakly correlated, the hydrodynamic regime can
occur precisely in that range of the spectrum which makes the main
contribution to the Landau formula. In this case, Eq.~(\ref{nl}) does not
hold. However, according to our direct estimation, even at a smallest
calculated density of $\bar n=1/256$, such a situation is not realized.

(4) In calculation of the local component, we set
$\varepsilon_p^T\approx\varepsilon_p^{T=0}=\varepsilon_p$ in Eqs.~(\ref{nl})
and (\ref{nep}). Indeed, at low temperatures, $T\ll T_q$, the Bose exponential
(\ref{nep}) leaves in (\ref{nl}) only the contribution from long-wavelength
scales, on which
$\varepsilon_p^T\approx c_s(T)p\approx c_s(0)p\approx\varepsilon_p$. At high
temperatures, ($T\sim T_q$, $T<T_q$), owing to the factor $(p^2/2)pdp$ in
Eq.~(\ref{nl}), long-wavelength scales do not contribute. On scales in the
range of shorter (microscopic) wavelengths, the spectrum calculated by the
Feynman formula is weakly affected by temperature variations. The latter fact
was revealed in the course of the Monte Carlo simulation of the pair
distribution function and the structure factor in liquid helium up to
temperatures so high (4.2 K) as the doubled temperature of the $\lambda$
point, $T_{\lambda}=2.1$ K \cite{b2103638}.

(5) The temperature of quasicondensation $T_q$ we formally calculate from the
condition $n_l(T_q)/n=0$, where $n_l(T)/n$ is given by Eq.~(\ref{nl}) with
$\varepsilon_p^T=\varepsilon_p$. This is an approximate estimate, since, in
the range of crossover at temperatures near $T_q$, the local superfluidity and
the local long-range order vanish, elementary excitations cease to form a
nearly ideal gas, and the quantum regime goes to the classical regime.

\section*{Acknowledgements}
This study was supported by the Russian Foundation for Basic Research, the
Swedish Research Council, VR, and the Swedish Foundation for Strategic
Research (SSR)


\begin{thebibliography}{99}
\bibitem{Butovring}
In some observations of structured luminescence rings \cite{b7500311},
excitons are weakly correlated. However, we suppose their density to be too
low, so, the transition to a collective state does not occur. Indeed, our
estimation shows (see Ref.~\cite{LKW}, and Sections \ref{measur}, \ref{oute},
\ref{outnl})) that, for the blueshift of the luminescence line observed in
\cite{b7500311} (being of the order of 0.5 meV), the exciton density is
$n=2.4\cdot10^9$ cm${}^{-2}$. So, the superfluid crossover temperature in a
trap with a size of $L=10\;\mu$ is $T_c^L=0.219\cdot2\pi\hbar^2n/m=0.13$ K,
which is much lower than the exciton temperature $T=1.6$ K.
\bibitem{nl=n}
The local superfluid exciton density $n_l$ (see Section \ref{outnl}) at the
superfluid transition temperature ($T=T_c$) is small ($n_l\ll n$) in 3D
systems and in quasi-2D systems with unquantized transverse motion. In
(quasi-)2D systems with quantized transverse motion, the inequality $n_l\ll n$
at $T=T_c$ holds true if the correlations are either weak or so strong that
the roton minimum is very deep. Therefore, the equality $n_l\approx n$ at
$T=T_c$ in 2D systems with quantized transverse motion holds only if
correlations are rather strong. In the case of intermediate correlations, the
fulfillment of the equality $n_l\approx n$ at $T=T_c$ is worse.
\bibitem{Butovblueshift}
In interacting exciton systems at $T=0$, the exciton resonance is redshifted
by a quantity related to $p_H$. This is due to that, at $T=0$, upon exciton
recombination, an elementary excitation is created from the ground state and
acquires a portion of the energy. This differs from the case of an ideal gas
at $T\ne0$, when an exciton with the energy $\hbar\Omega+p_H^2/2m$ recombines
and transfers all this energy to an emitted photon causing a blueshift. The
latter scenario was observed in \cite{BH}.
\end{thebibliography}
\end{document}